\documentstyle[aps,twocolumn]{revtex}

\newcommand{\sa}[1]{\scriptscriptstyle{#1}}

\begin{document}

\title{The Quantum State of a Propagating Laser Field}

\author{S.~J. van Enk and Christopher A. Fuchs}
\address{Bell Labs, Lucent Technologies, 600-700 Mountain Ave,
Murray Hill, NJ 07974, U.S.A.}
\date{27 November 2001}

\maketitle

\begin{abstract}
Optical implementations of quantum communication protocols typically
involve laser fields.  However, the standard description of the
quantum state of a laser field is surprisingly insufficient to
understand the quantum nature of such implementations.  In this
paper, we give a quantum information-theoretic description of a
propagating continuous-wave laser field and reinterpret various
quantum-optical experiments in light of this.  A timely example is
found in a recent controversy about the quantum teleportation of
continuous variables.  We show that contrary to the claims of T.
Rudolph and B. C. Sanders [Phys.\ Rev.\ Lett.\ {\bf 87}, 077903
(2001)], a conventional laser can be used for quantum teleportation
with continuous variables and for generating continuous-variable
quantum entanglement.  Furthermore, we show that optical coherent
states do play a privileged role in the description of propagating
laser fields even though they cannot be ascribed such a role for the
intracavity field.
\end{abstract}
\medskip

\section{Introduction}

What is the quantum state of a laser field? According to textbook
laser theory---see for example Chapter 17 in Ref.~\cite{lamb} or
Chapter 12 in Ref.~\cite{walls}---the quantum state of the field
{\em inside\/} the laser cavity in a steady state is a mixed state
diagonal in the photon-number basis, with Poissonian number
statistics:
\begin{equation}\label{yumyum}
\rho_{|\alpha|} = e^{-|\alpha|^2}\sum_n
\frac{|\alpha|^{2n}}{n!}|n\rangle\langle n|\;.
\end{equation}
Such a state does not have a well-defined phase. Yet, laser fields
are routinely used to define a phase standard. Moreover, many, if
not all, standard optics experiments seem to be consistent with the
assumption that the laser field is in a pure coherent state.

The common explanation for how this comes about seems most often to
rest on the mathematical identity
\begin{equation}\label{identity}
e^{-|\alpha|^2}\sum_n \frac{|\alpha|^{2n}}{n!}|n\rangle\langle n|
=\int \frac{{\rm d}\varphi}{2\pi} |\alpha
e^{i\varphi}\rangle\langle\alpha e^{i\varphi}|\;,
\end{equation}
where the right-hand side denotes an integral over all coherent
states $|\alpha e^{i\varphi}\rangle$ with amplitude $|\alpha|$.
Through this identity one might think at first sight that one has the
right to think of laser light as a mixture of coherent states:  ``The
ideal laser field really is in some coherent state $|\alpha
e^{i\varphi}\rangle$, we just do not know which one.''  (See for
instance, Ref.~\cite[pp.~15, 38]{walls}.)  However, this kind of
thinking already carries the seed of its own demise.  For if one can
use this argument to imagine the existence of an unknown coherent
state for describing the true state of the laser field, then one can
use it just as well to imagine the existence of an unknown number
state for the same task.  To think otherwise is to commit the
so-called {\it preferred ensemble fallacy\/} (PEF), a move that has
no justification within standard quantum mechanics. (For a general
discussion of the PEF, see Ref.~\cite{CavesFuchs96}. For some other
examples of the havoc it can cause in quantum information science,
see Refs.~\cite{Braunstein99,Kok00}.)

What is the solution to this conundrum?  Of course, one can argue
that any experiment whose outcome does not depend on the absolute
phase $\varphi$ cannot distinguish between a pure-state $|\alpha
e^{i\varphi}\rangle$ and a mixed-state $\rho_{|\alpha|}$. However,
this observation is not general enough to explain many of the
results of present-day optics experiments. For instance, it does not
yet explain why a phase measurement between two independent laser
beams will give the same result as a subsequent phase measurement on
more light emanating from the same lasers, or how coherent
manipulations of atoms and ions are in fact possible even though no
coherent superpositions of atomic states are produced. It also may
not be clear what is meant by phase diffusion in the context of
laser theory---see for example Chapter 20 of Ref.~\cite{lamb}---if
the phase of the laser cavity field is totally random.

M\o lmer was perhaps the first to address the apparent contradiction
between the two different descriptions of a laser field in
Ref.~\cite{molmer}. In particular, he studied a standard measurement
of the phase between two independent light beams emanating from
cavities initially in pure {\em number states}, so that neither
state has a well-defined phase by itself nor relative to the other.
He showed that indeed a definite phase difference will be measured;
it is just that that value will be random from experiment to
experiment. Nevertheless, if one keeps monitoring the phase
difference between two such light beams, the randomly established
phase value will persist.

In this paper, we contribute to the furtherance of this discussion by
diverting attention away from the intracavity field and refocusing it
on the field after it leaks out of the cavity, i.e., on the {\em
propagating\/} laser beam.  In particular, we reexamine the question
of the quantum state of a propagating laser field from a
quantum-information theoretic perspective. Using this description,
the solution to the conundrum above becomes clear.  Moreover, it also
allows us to answer an important question raised in
Ref.~\cite{rudolph}. There it is concluded that teleportation with
continuous variables is not possible with a mixed state of the form
(\ref{yumyum}), but requires a true coherent state. The main reason
for their conclusion is the contention that a mixture of two-mode
squeezed states produced by a laser in a mixed state does not
contain any quantum entanglement. This is an important observation.
In fact, this is a splendid example of why Eq.~(\ref{identity}) does
not capture the complete essence of most experiments with laser
light. Our formulation clarifies why the coherent state plays a
privileged and unique role in the description of propagating laser
fields, and how a conventional laser can produce quantum
entanglement, even if it cannot actually produce a two-mode squeezed
state.

The plan of the remainder of the paper is as follows.  In Section
II, we use the input-output formalism of Refs.~\cite{walls,collett}
to derive a multi-mode description of an ideal laser beam.  We then
use the quantum de Finetti theorem \cite{hudson,caves} to show in
what sense the expression just derived is a unique one.  In Section
III, we apply the results of Section II to elucidate several
examples:  measurement of the relative phase between two independent
lasers, the coherent excitation of a two-level atom, the production
and detection of squeezed states, the production of two-mode
squeezed states, and finally quantum teleportation of continuous
variables.  In Section IV, we give some concluding remarks.

For all of the examples addressed in Section III the description of
an {\em ideal\/} noiseless laser is sufficient for understanding the
point of principle. It is nevertheless interesting and important to
describe phase diffusion and its effects on the quantum state of a
laser beam.  To that end, we further give a discussion of the
phase-diffusing laser in light of the present considerations in
Section IIC.  Remarks are also made throughout Section III concerning
the effect of phase diffusion for the phenomenon at hand.

\section{The Quantum State of a Propagating Laser Beam}

We are interested in calculating the quantum state of the light
field of a continuous-wave (CW) laser. We model the laser as a
one-sided cavity driven by a constant force (a voltage or an
external field) far above threshold. We first consider an imaginary
case where the field inside the cavity is in a coherent state and
calculate the quantum state of the field outside the laser cavity.
Then, using the identity (\ref{identity}), we use the linearity of
quantum mechanics to derive the true state of a laser field outside
of the cavity. Subsequently we imagine that we partition the light
beam into packages of equal length (or duration) and rewrite the
result in terms of the quantum states of the individual packages. We
then compare that result with the general form of the quantum state
of an ensemble produced by a source that emits unknown but identical
states. Next we consider a more realistic model of a laser and
include phase diffusion.

\subsection{Input-Output Relations}

We employ standard input-output theory \cite{walls,collett} to
connect the quantum field inside a laser cavity to its output field.
First, we separate the field modes into two parts. A single-mode
annihilation operator $a$ describes the field with frequency
$\omega_0$ inside the cavity; continuous-mode operators $b(\omega)$
describe modes with frequency $\omega$ outside the cavity. This
separation is an approximation. It is valid for (1) high-finesse
cavities (i.e., the cavity decay rate $\kappa$ must be much smaller
than the distance between adjacent resonant frequencies, $\pi c/L$,
with $L$ the length of the cavity) (2) on time scales much longer
than a cavity round trip time $L/c$ and (3) for frequencies near
resonance (i.e., for frequencies $\omega$ within a few $\kappa$ from
the relevant resonance frequency, $\omega_0$, in the
problem)~\cite{vogel}.

We define input and output operators by
\begin{eqnarray}
a_{{\rm in}}(t) &=& \frac{-1}{\sqrt{2\pi}}\int{\rm d}
\omega e^{-i\omega (t-t_0)}b_0(\omega),\nonumber\\
a_{{\rm out}}(t) &=& \frac{1}{\sqrt{2\pi}}\int{\rm d}\omega
e^{-i\omega (t-t_1)}b_1(\omega),
\end{eqnarray}
where $t_0\rightarrow -\infty$ is a time in the far past and
$t_1\rightarrow\infty$ is a time in the far future. The operators
$b_0(\omega)$ and $b_1(\omega)$ are defined to be the Heisenberg
operators $b(\omega)$ at times $t=t_0$ and $t=t_1$, respectively.
The input and output operators $a_{{\rm in,out}}(t)$ are
not themselves Heisenberg operators. They do satisfy, by
construction, bosonic commutation relations for free-field
continuous-mode operators,
\begin{eqnarray}
[a_{{\rm in,out}}(t),a^{\dagger}_{{\rm in,out}}(t')]=\delta(t-t').
\end{eqnarray}
The integrations over frequencies $\omega$ extend from $-\infty$ to
$+\infty$. This, obviously, involves an approximation. Namely
$\omega$ is in fact defined as the frequency relative to the
relevant resonance frequency we are interested in (i.e., we moved to
a rotating frame), and thus ranges in principle from $-\omega_0$ to
$\infty$. Since $\omega_0$ is usually by far the largest frequency
in the problem, it is a good approximation to extend the limit down
to $-\infty$. This is valid as long as we consider times scales much
larger than $1/\omega_0$.

Since nonlinear effects are typically very small, the interaction
between the cavity mode and the outside modes can be well
approximated by a linear interaction. Using the fact that the
coupling (the cavity decay rate) is more or less constant over the
relevant range of frequencies, one may choose the coupling to be
constant over the entire range of frequencies to good approximation.
The resulting input-output relation
\begin{equation}\label{boundary}
a_{{\rm in}}(t)+a_{{\rm out}}(t)=\sqrt{\kappa}a(t),
\end{equation}
with $\kappa$ the decay rate of the cavity takes then a simple form,
which can in fact be interpreted as a boundary condition on the
electric field. It is important to note that this equation is valid
irrespective of the internal dynamics of the cavity mode.

\subsection{The Noiseless Laser}\label{temp}
In this subsection we neglect dissipation and noise due to, e.g.,
spontaneous emission in the atomic laser medium. In this case the
field inside the laser cavity is well approximated when the laser
operates far above threshold by a state of the form (\ref{identity}).

When the input field is the vacuum and the field inside the cavity is
for the moment assumed to be a coherent state $|\alpha
e^{i\phi}\rangle$ with a known phase, then according to
(\ref{boundary})  the output field is an eigenstate of $a_{{\rm
out}}$ with eigenvalue $\beta(t)\equiv \sqrt{\kappa}\alpha
e^{i\phi}$. Such a state is a continuous-mode coherent state
\cite{loudon} and can be written in the Schr\"odinger picture as
\begin{equation}\label{coco}
|\{\beta(t)\}\rangle\equiv\exp\!\left(\int{\rm d}\omega [
\beta(\omega) b^{\dagger}(\omega) -
\beta^*(\omega)b(\omega)]\right)|{\rm vac}\rangle,
\end{equation}
with $|{\rm vac}\rangle$ is the vacuum state and $\beta(\omega)$ is
the Fourier transform of $\beta(t)$. In the idealized case without
noise one simply has a stationary laser beam with fixed frequency
$\omega_0$. A continuous-mode coherent state can be described
alternatively as an infinite tensor product of discrete-mode
coherent states \cite{loudon}. Define a complete set of functions
$\{\Phi_i(t)\}$ satisfying the following orthogonality and
completeness relations,
\begin{equation}\label{complete1}
\int {\rm d}\tau \Phi_i(\tau) \Phi_j^*(\tau) = \delta_{ij},
\end{equation}
and
\begin{equation}\label{complete2}
\sum_i \Phi_i(t)\Phi_i^*(t') = \delta(t-t').
\end{equation}
We may then define annihilation operators $c_i$ (satisfying the
correct bosonic commutation relations for discrete operators)
according to
\begin{equation}
c_i=\int {\rm d}t \Phi^*_i(t) a_{{\rm out}}(t).
\end{equation}
An eigenstate of $a_{{\rm out}}(t)$ with eigenvalue $\beta(t)$ is
also an eigenstate of $c_i$ with eigenvalue
\begin{equation}
\alpha_i=\int {\rm d}t\Phi^*_i(t) \beta(t).
\end{equation}
We now apply this formalism to describe laser light as a sequence of
packets of light, each with the same duration $T$. We thus define a
set of functions $\{ \Psi_n(t) \}$ by
\begin{equation}\label{Psin}
\Psi_n(t) = \left\{
\begin{array}{cl} \displaystyle \frac{1}{\sqrt{T}} & \,\,\, {\rm
for}\,\,\, \displaystyle \left|t-\frac{z_0}{c}-nT\right|<\frac{T}{2},\\
& \\
 0 & \,\,\,
{\rm otherwise.}
\end{array}\right. 
\end{equation}
The label $z_0$ refers to an arbitrarily chosen reference position
$z_0$ relative to which we partitioned the light beam into equal
pieces of length $cT$ (see Figure~1). This set of functions is orthogonal and can
be extended to form a complete set satisfying (\ref{complete1}) and
(\ref{complete2}). For a CW laser described by
$\beta(t)=\sqrt{\kappa}\alpha e^{i\phi}$ we see that each part $n$
of the light beam is in the same coherent state with eigenvalue
\begin{equation}
\alpha_n=\sqrt{\kappa T} \alpha e^{i\phi}\equiv \alpha_0,
\end{equation}
corresponding to the modes described by (\ref{Psin}), and
$\alpha_i=0$ for all other modes. The duration $T$ must be much
larger than both $1/\omega_0$ and $L/c$ but can be arbitrarily chosen
otherwise.

Now assuming that the field inside the laser cavity is in fact a
mixture $\rho_{|\alpha|}$, the quantum state of a sequence of $N$
parts corresponding to the set $\{ \Psi_n\}$ is thus
\begin{equation}\label{ensemble}
\tilde{\rho}_N =\int \frac{{\rm d}\varphi}{2\pi}\big( |\alpha_0
e^{i\varphi}\rangle \langle \alpha_0 e^{i\varphi}|\big)^{\otimes
N}\;,
\end{equation}
where the integrand signifies an $N$-fold tensor product over the
separate packets.

\subsection{The Phase-Diffusing Laser}
Equation (\ref{ensemble}) is the quantum state of an ideal
propagating laser field. Now let us consider a more realistic laser
and take into account the effects of noise and dissipation. We
present here only the main ideas and results from standard laser
theory and use those to connect the field inside the laser cavity to
the outside field we are interested in. For technical details we
refer the reader to Ref.~\cite{lamb}.

A first consequence of the presence of noise is that the state of the
laser cavity field is not a steady state. Instead, there is in
general a decay towards a steady state. The diagonal matrix elements
(in the number-state basis) of the density matrix of the field decay
towards a Poissonian distribution, the off-diagonal matrix elements
decay to zero. The steady state at any time is, therefore, still a
mixture of coherent states with random phases. The rate at which
that steady state is reached is proportional to the average number
of photons inside the cavity and inversely proportional to the
quality factor of the cavity. Far above threshold and for
high-quality cavities we may then approximate the state of the
cavity field by its steady state value, as long as we consider time
scales longer than the time needed to reach equilibrium.

For the purpose of finding the quantum state of the field outside the
laser the most convenient representation is the Heisenberg picture,
so that we can apply the input-output relation (\ref{boundary}).
Chapter 20 of Ref.~\cite{lamb} derives an equation for the operator
$a(t)$. When the atomic operators have been eliminated the resulting
equation for $a(t)$ turns out to be nonlinear (it contains a term
proportional to $\langle a^{\dagger}(t)a(t)\rangle a(t)$) and to
contain fluctuating noise terms. This equation is hard to solve in
general but far above threshold one may go to the ``classical''
limit and write down an equation for the expectation value of $a(t)$
while using a decorrelation approximation that replaces $\langle
a^{\dagger}(t)a(t)\rangle$ with $|\langle a(t)\rangle|^2$. This is
basically the same approximation as the steady-state approximation
mentioned above, except that it implicitly assumes, for the moment,
that the steady state is a coherent state, rather than a mixture of
coherent states. The intermediate result then gives us a coherent
state whose amplitude and phase fluctuate randomly with time.
However, the phase fluctuations are typically much larger than the
amplitude fluctuations and this leads to the concept of a
phase-diffusing laser. The amplitude of the field in the cavity is
approximated to be $\alpha e^{i\phi}e^{i\eta(t)}$ with $\eta(t)$ a
random Gaussian variable with correlations given by
\begin{equation}
\langle a^{\dagger}(t) a(0)\rangle=\alpha^2 \exp(-Dt),
\end{equation}
with $D$ the phase diffusion constant (determined by spontaneous
emission rate of the atomic medium, the quality factor of the laser
cavity and the average number of photons). According to
(\ref{boundary}) the outside field is still a continuous-mode
coherent state with amplitude $\beta(t)=\sqrt{\kappa}\alpha
e^{i\phi} e^{i\eta(t)}$, but it is no longer monochromatic since
$\beta$ is not constant. Indeed, the laser linewidth is given by
$D$. Nevertheless, we can still use the same complete set of
functions $\Phi_i(t)$ of which the functions $\Psi_n(t)$ of
Eq.~(\ref{Psin}) are a subset. There are two types of modifications
to the result (\ref{ensemble}). First, the modes corresponding to
$\Psi_n(t)$ are no longer the only ones with a nonzero amplitude,
and second, the amplitudes $\alpha_n$,
\begin{equation}
\alpha_n=\int {\rm d}t\Psi^*_n(t) \beta(t),
\end{equation}
are no longer constant in magnitude, nor do they all have the same
phase. The absolute value of the amplitude is in fact reduced since
\begin{equation}\label{eta}
|\alpha_n|=\left|\int_{z_0/c+(n-1/2)T}^{z_0/c+(n+1/2)T}  {\rm d}t
e^{i\eta(t)}\right|\frac{\alpha}{\sqrt{T}},
\end{equation}
and the integral over time is less than $T$. This is consistent with
the fact that other modes carry a finite amount of light as well.

If we choose $T$ to be much smaller than a diffusion time $1/D$ (but
still larger than an optical period and a cavity roundtrip time),
then the amplitudes $|\alpha_n|$ are almost equal to what they would
be in the absence of noise since the integral appearing in
(\ref{eta}) is almost equal to $T$. Hence, in that case the quantum
state of the laser is well approximated by
\begin{eqnarray}\label{ensemblen}
\tilde{\rho}_N &=&\int {\rm d}\epsilon_1\int{\rm d}\epsilon_2
\cdots \int{\rm d}\epsilon_k  \int \frac{{\rm d}\varphi}{2\pi} |\alpha_0
e^{i\varphi}\rangle \langle \alpha_0 e^{i\varphi}|\nonumber\\
&&\otimes P(\epsilon_1)|\alpha_0
e^{i\varphi+i\epsilon_1}\rangle \langle \alpha_0
e^{i\varphi+i\epsilon_1}|
\nonumber\\
&& \otimes P(\epsilon_2)
|\alpha_0 e^{i\varphi+i\epsilon_1+i\epsilon_2}\rangle \langle
\alpha_0 e^{i\varphi+i\epsilon_1+i\epsilon_2}|\;
\cdots \nonumber\\
&& \otimes P(\epsilon_k)
|\alpha_0 e^{i\varphi+i\sum_{k=1}^N\epsilon_k}\rangle \langle
\alpha_0 e^{i\varphi+i\sum_{k=1}^N\epsilon_k}|,
\end{eqnarray}
for the modes corresponding to $\Psi_n$ with the remaining modes
almost empty. Here we introduced the random variables $\epsilon_k$ by
\begin{equation}
\epsilon_k={\rm Im}\!\left[\log
\int_{z_0/c+(k-1/2)T}^{z_0/c+(k+1/2)T} {\rm d}t e^{i\eta(t)}\right],
\end{equation}
with average zero and variance $\sigma=\sqrt{2DT}\ll 1$, and the corresponding probability distribution
\begin{equation}
P(\epsilon)=\frac{1}{\sqrt{2\pi\sigma^2}}\exp\left(-\frac{\epsilon^2}{2\sigma^2}\right).
\end{equation}

\subsection{The Quantum de Finetti Theorem}
The result (\ref{ensemble}) for an ideal propagating laser field
displays an apparent privileged role for coherent states in
describing a propagating laser field: Although the quantum state
inside the laser is a mixed state diagonal in the number-state
basis, the quantum state of the output is not equal to a product of
mixed states $(\rho_{|\alpha_0|})^{\otimes N}$ (as it would be for a
pulsed laser). Rather it is a mixture of $N$ copies of a coherent
state, each copy with the same ``unknown'' phase. The real question
is, is this the only such description?  The last thing we would want
to do is commit the {\it preferred ensemble fallacy\/} (PEF) that
Rudolph and Sanders~\cite{rudolph} rightly warn against.  In other
words, in analogy to Eq.~(\ref{identity}), how do we know that there
may not be some other way of representing the density operator in
Eq.~(\ref{ensemble}), say by
\begin{equation}\label{ensembleprime}
\tilde{\rho}_N =\int d\Omega_\psi \big( |\psi\rangle \langle
\psi|\big)^{\otimes N}\;,
\end{equation}
where the $|\psi\rangle$ represent a completely different set of
states than the coherent states and $d\Omega_\psi$ represents a
measure on that set. The answer to this question lies in the quantum
de Finetti representation theorem~\cite{hudson,caves}.

Consider a source that sequentially introduces an {\it infinite\/}
set of quantum systems, the first $N$ of which are described by a
density operator $\tilde{\rho}_N$.  We shall make two requirements
of this {\it sequence\/} of density operators.  First, they should
all be compatible in the sense that $\tilde{\rho}_N$ can be derived
from $\tilde{\rho}_{N+1}$ by performing a partial trace over the
Hilbert space of the $(N+1)$'th system.  And second, for each $N$,
$\tilde{\rho}_N$ should have the property that interchanging any two
of the systems will not change the joint probability distribution
for the outcomes of measurements on any of the individuals---that is
to say, the density operator $\tilde{\rho}_N$ should remain
invariant under a permutation of the systems it describes.

The quantum de Finetti representation theorem~\cite{hudson,caves}
specifies that---with these assumptions alone---the quantum state of
any $N$ systems from such a source can {\it necessarily\/} be
written in the form
\begin{equation}\label{Finetti}
\tilde{\rho}_N=\int {\rm d}\rho P(\rho) \rho^{\otimes N}\;,
\end{equation}
where $P(\rho)$ is a probability distribution over the density
operators and ${\rm d}\rho$ is a measure on that space. Most
importantly for the considerations here, this representation is {\it
unique\/} up the behavior of $P(\rho)$ on a set of measure zero.

The meaning of this result in the present context is manifest: To the
extent that one believes that a laser beam can be chopped into equal
pieces and rearranged without affect to one's experiments---that is,
that the beam is {\it stationary}---the representation in
(\ref{ensemble}) is the only possibility of the form
Eq.~(\ref{ensembleprime}).  That is to say, one can always act
\underline{{\it as if\/}} the temporal modes of a propagating laser
beam are all in the same fixed but {\it unknown\/} coherent state.
There are no other quantum states that will fit this bill.

Indeed, contemplate performing a set of measurements on the
individual systems emanating from such a source as above. As the data
accumulates, the probability distribution $P(\rho)$ in
(\ref{Finetti}) should be updated according to standard Bayesian
rules after the acquisition of that information~\cite{schack}.
Specifically, if measurements on $K$ systems yield results $D_K$,
then the state of additional systems is constructed as in
Eq.~(\ref{Finetti}), but using an updated probability distribution
given by
\begin{equation}
P(\rho|D_K)={P(D_K|\rho)P(\rho)\over P(D_K)}\;.
\label{QBayes}
\end{equation}
Here $P(D_K|\rho)$ is the probability to obtain the measurement
results $D_K$, given the state $\rho^{\otimes K}$ for the $K$
measured systems, and
\begin{equation}
P(D_K)=\int P(D_K|\rho)\,P(\rho)\,d\rho
\end{equation}
is the unconditional probability for the measurement results.

For a sufficiently informative set of measurements---namely a set of
measurements whose eigenspaces span the whole linear vector space of
operators over the initial Hilbert space---as $K$ becomes large, the
updated probability $P(\rho|D_K)$ becomes highly peaked on a
particular state $\rho_{D_K}$ dictated by the measurement results,
regardless of the prior probability $P(\rho)$, as long as $P(\rho)$
is nonzero in a neighborhood of $\rho_{D_K}$.   In other words, the
measurement results essentially collapse the original mixed state to
a new one in which any number $M$ of additional systems are assigned
the product state $\rho_{D_K}^{\otimes M}$, i.e.,
\begin{equation}
\int P(\rho|D_K)\,\rho^{\otimes M}\,d\rho \quad{\longrightarrow}\quad
\rho_{D_K}^{\otimes M}
\label{HannibalLecter}
\end{equation}
for $K$ sufficiently large.

Comparing the state of a propagating laser field (\ref{ensemble})
with the general form (\ref{Finetti}) we see that a {\em complete\/}
set of measurements on part of the light emanating from the laser
will reduce the quantum state of the rest of the light to a pure
state.  But most importantly, this pure state will be a {\em
coherent state}---there is no way to make it a number state; there is
no way to make it a squeezed state or any other kind of exotic
state.  In this sense, the coherent states play a privileged role in
the description of laser light \cite{JulioThing}.

It is true that standard optics experiments have not yet featured
such complete measurements. For instance, a complete set for the case
at hand would be a measurement of amplitude and absolute phase.
However, recent developments \cite{jones} may make it possible to
compare the phase of an optical light beam directly to the phase of
a microwave field. Using this technique the only further measurement
required for a complete measurement is a measurement of the absolute
phase of the microwave field, which is possible electronically. This
measurement would create an optical coherent state from a standard
laser source for the first time. But as we will show in the next
section, such a measurement does not even need to be performed for
many applications.

\subsection{Beamsplitters}

In Section \ref{temp} we constructed a set of {\em temporal\/} modes
and used those to write down the quantum state of a propagating laser
field. Alternatively, one may use {\em spatial\/} modes. In
particular, we may imagine dividing a laser beam into an arbitrary
number, $N$, of spatially different pieces by using beamsplitters.
In general, the action of a beamsplitter with reflection and
transmission coefficients $r,t$ on a coherent state is given by
\begin{equation}\label{bs}
|\alpha e^{i\phi}\rangle|0\rangle\mapsto
|r\alpha e^{i\phi}\rangle|t\alpha e^{i\phi}\rangle,
\end{equation}
where the notation indicates that both input and output of a
beamsplitter consist of two modes, and where here one input mode is
in the vacuum state. Now given one discrete mode in a coherent state
we may indeed apply the transformation (\ref{bs}) multiple times to
obtain a state of the form (\ref{ensemble}). This method may seem
simpler than that used in Section \ref{temp} to derive
(\ref{ensemble}) but it has two disadvantages. First, one still has
to justify the discrete mode one started with. Indeed, one choice
would be to choose a particular {\em temporal\/} mode. Second, once
one has chosen a beamsplitter setup that divides the laser into $N$
spatial modes, each carrying the same amount of light, one can no
longer extend the set to $N+1$ spatial modes without changing the
amplitudes of the original $N$ modes. The de Finetti theorem could,
therefore, not be applied to a sequence of quantum states so
constructed. Of course, to extend a set of {\em temporal\/} modes by
a further temporal one is easy:  One just waits a little longer.

\section{Mixed-State Description of Optical Experiments}
Let us now describe a few typical optical experiments using
(\ref{ensemble}) for a proper description of the quantum state of a
laser. This is sufficient for our purposes but we also
discuss how the effects of phase diffusion modify the description.

\subsection{Phase Measurement for Independent Lasers}
M\o lmer in \cite{molmer} showed that the detection of a phase
difference between two (independent) light beams need not imply that
there is a well-defined phase difference before the measurement. In
particular, he showed that for light emanating from two cavities
whose fields are initially in number states (whose phase is
completely random), the standard setup to measure phase will indeed
find a stable phase difference (though the value of this phase will
be random and different from experiment to experiment). Within one
experiment, it takes just a few (about three) photon detections
\cite{molmer} to settle on a particular value of the phase
difference, after which the counting rates of the detectors remain
consistent with that initial phase difference. In other words, the
standard phase measurement acts almost like a perfect von Neumann
measurement; the measurement will produce an eigenvalue of the
corresponding observable and the state after the measurement can be
described by an eigenstate of the measured variable.

Generalizing this observation to continuously pumped CW lasers leads
to the following simple description. Initially we have two
independent laser beams $A$ and $B$ whose joint quantum state is
described by
\begin{eqnarray}\label{2}
\tilde{\rho}_{2N} &=& \int \frac{{\rm d}\varphi_A}{2\pi}\big(
|\alpha_{\sa{A}} e^{i\varphi_{\sa{A}}}\rangle \langle \alpha_{\sa{A}}
e^{i\varphi_{\sa{A}}}|\big)^{\otimes N}
\nonumber\\
&& \otimes \int \frac{{\rm d}\varphi_{\sa{B}}}{2\pi}
\big( |\alpha_{\sa{B}} e^{i\varphi_{\sa{B}}}\rangle \langle
\alpha_{\sa{B}} e^{i\varphi_{\sa{B}}}|\big)^{\otimes N}
\end{eqnarray}
if we divide each laser beam into $N$ packages of constant duration.
If the first package of each beam is used to measure a phase
difference then the state of the rest of the light beams will be
reduced to
\begin{eqnarray}\label{2'}
\tilde{\rho}_{2N-2} &=& \int \frac{{\rm d}\varphi_A}{2\pi}\big(
|\alpha_{\sa{A}} e^{i\varphi_{\sa{A}}}\rangle \langle \alpha_{\sa{A}}
e^{i\varphi_{\sa{A}}}|
\big)^{\otimes (N-1)}\nonumber\\
&& \otimes \big( |\alpha_{\sa{B}} e^{i(\phi_0+\varphi_{\sa{A}})}\rangle
\langle \alpha_{\sa{B}} e^{i(\phi_0+\varphi_{\sa{A}})}|\big)^{\otimes
(N-1)},
\end{eqnarray}
where we assumed the outcome of the phase measurement was $\phi_0$
and approximated the measurement to be sharp. The state (\ref{2'})
has the property that a subsequent measurement of the phase
difference will reproduce the value $\phi_0$: This is a kind of
``phase-locking without phase.'' Note this would certainly not be
the case if the quantum state of a laser were a product of identical
mixed states of the form $(\rho_{|\alpha|})^{\otimes N}$. Also note
that in the number-state basis such properties are hard to
understand.

For a phase-diffusing laser, (\ref{ensemblen}) shows that
measurements on adjacent parts of the laser beam will give
approximately the same value for phase, whereas measurements on
parts that are further apart than the diffusion time will give
random results.

\subsection{Coherent Excitation of a Two-Level Atom}\label{atom}

In many experiments atomic coherence has been demonstrated:
Superpositions of atomic states, degenerate or nondegenerate, have
been supposedly created by using lasers. But how can one create such
a coherent superposition if the laser field apparently is not
coherent? Let us describe a typical experiment. If a laser would
produce a coherent state $|\alpha\rangle$ it could be used to create
an equal superposition of ground and excited states of a two-level
atom by applying a $\pi/2$ pulse.  In the limit of a large
coherent-state amplitude the laser field would not become entangled
with the atom\cite{enkkimble} and one can write for the process
taking place\cite{noteD}
\begin{equation}
|\alpha\rangle\otimes |g\rangle \mapsto |\alpha\rangle \otimes
(|g\rangle+|e\rangle)/\sqrt{2}
\end{equation}
with $|g,e\rangle$ the atomic ground and excited states.
However, taking into account the actual state of a laser field we have
\begin{eqnarray}
&&\int \frac{{\rm d}\varphi}{2\pi}\big( |\alpha
e^{i\varphi}\rangle \langle \alpha e^{i\varphi}|\big)^{\otimes
N}\otimes \Pi_g \nonumber\\ &\hookrightarrow& \int \frac{{\rm
d}\varphi}{2\pi}\big( |\alpha e^{i\varphi}\rangle \langle \alpha
e^{i\varphi}|\big)^{\otimes N} \otimes \Pi_\phi\;,
\end{eqnarray}
where $\Pi_g$ is the projector onto the ground state $|g\rangle$ and
\begin{equation}
\Pi_\phi=\frac{1}{2}\big(|g\rangle+e^{i\phi}|e\rangle\big)\big(\langle
g|+e^{-i\phi}\langle e|\big).
\end{equation}
The ``phase'' of the atomic superposition is equal to the ``phase''
of the coherent state. If we would trace out the laser field the
resulting atomic density matrix would be completely mixed, an equal
mixture of ground and excited states. However, in every experiment
exploiting atomic coherence it is the same laser (or one that has
been ``phase-locked'' to the same laser) that is used to perform a
measurement on the atom. For instance, one applies another $\pi/2$
pulse to take the atom to the excited state and subsequently
measures the atomic population in ground and excited states. But
this works just as well with a laser in the mixed state since the
{\em overall\/} process is independent of the phase of the laser.
That is, by applying the second $\pi/2$ pulse one gets
\begin{eqnarray}
&&\int \frac{{\rm d}\varphi}{2\pi}\big( |\alpha
e^{i\varphi}\rangle \langle \alpha e^{i\varphi}|\big)^{\otimes
N}\otimes \Pi_\phi
\nonumber\\ &\hookrightarrow& \int \frac{{\rm d}\varphi}{2\pi}\big(
|\alpha e^{i\varphi}\rangle \langle \alpha
e^{i\varphi}|\big)^{\otimes N} \otimes \Pi_e,
\end{eqnarray}
and the probability to detect the atom in the state $|e\rangle$ is
unity, whereas this probability would be 1/2 if uncorrelated laser
beams would have been used.

If one considers this same experiment in the number-state basis the
fact that an incoherent mixture of excited and ground states is
transformed into a pure excited state is rather miraculous and seems
to depend on very special correlations between the original laser
pulse and the measurement pulse \cite{bea}.

\subsection{Production and Detection of Squeezed States} A squeezed
state may be produced with the help of a nonlinear process described
by an interaction Hamiltonian
\begin{equation}
H_I=\chi[a^{\dagger}b^2 + b^{\dagger 2}a],
\end{equation}
where $a$ and $b$ are annihilation operators of single modes inside
an optical resonator with frequencies $\omega_0$ and $\omega_0/2$,
respectively, and $\chi$ is proportional to the second-order
nonlinearity $\chi^{(2)}$ of the nonlinear medium placed inside the
cavity. Pump photons at frequency $\omega_0$ can be downconverted to
pairs of photons of frequency $\omega_0/2$. The resulting quantum
state of the downconverted photons may display nonclassical
two-photon correlations. If the pump field is in a coherent state
and the mode $b$ is initially in the vacuum state, then the state
produced is a squeezed vacuum. On the other hand, if the pump field
is in a mixed state diagonal in the number state basis, then the
resulting state of mode $b$ is also diagonal in the number state
basis, since the interaction $H_I$ preserves the number operator
$N=2a^{\dagger}a+b^{\dagger}b$.

As a consequence, if we write $|S_{\alpha}(\varphi)\rangle$ for a
squeezed vacuum state produced by a laser in a coherent state with
amplitude $\alpha e^{i\varphi}$ with $\alpha$ real, then the mixed
state
\begin{equation}\label{sq}
\int \frac{{\rm d}\varphi}{2\pi}
|S_{\alpha}(\varphi)\rangle \langle S_{\alpha}(\varphi)|
\end{equation}
is not a squeezed state and does not display any nonclassical
features, as this state too is a mixed state diagonal in the photon
number state basis. (Similarly, in a simple picture the degrading effect of
phase diffusion on squeezing can be understood by considering a
mixture of squeezed states with phases drawn from a Gaussian
probability distribution. A more detailed discussion is given in
\cite{drummond}.) Yet, the state that is actually obtained in an
experiment is of the form
\begin{equation}\label{sq2}
\int \frac{{\rm d}\varphi}{2\pi} \big(|S_{\alpha}(\varphi)\rangle
\langle S_{\alpha}(\varphi)|\big)^{\otimes M} \otimes \big( |\alpha
e^{i\varphi}\rangle \langle \alpha e^{i\varphi}|\big)^{\otimes
(N-M)},
\end{equation}
where we assumed that $M$ light packets traversed the nonlinear
medium and $N-M$ did not. Note that tracing out the unsqueezed part leaves a 
residue of de Finetti form: What remains can be viewed as a mixture of identical copies
of some unknown squeezed state. Furthermore, using the exchangeability of that
density operator, the quantum de Finetti theorem implies this expansion is unique.
A complete tomographic measurement on some of those copies will reduce the quantum state of
the remaining copies to a simple tensor product of squeezed states.

The state (\ref{sq2}) will display
nonclassical correlations between the squeezed mode(s) and the
remaining laser light. In fact, those correlations do not depend on
the value of $\varphi$, and, therefore, are the same as those that
would be measured if one had a coherent state. The distinction
between properties of a state like (\ref{sq}) and the same state but
with the correlations with the laser light included, becomes more
pronounced when we consider a two-mode squeezed state.

\subsection{Production of Two-Mode Squeezed States}
A two-mode squeezed state can be generated by splitting two squeezed
states on a 50-50 beamsplitter. The resulting state of the two
output ports is an entangled state. Denote a two-mode squeezed state
generated from a coherent state with amplitude $\alpha e^{i\varphi}$
by $|T^{\sa{AB}}_\alpha(\varphi)\rangle$, where the superscripts
$A,B$ refer to two distinct modes located in different laboratories,
say Alice's and Bob's. As shown in \cite{rudolph}, the state
\begin{equation}\label{t}
\int \frac{{\rm d}\varphi}{2\pi} |T^{\sa{AB}}_\alpha(\varphi)
\rangle\langle T^{\sa{AB}}_\alpha(\varphi)|
\end{equation}
contains no entanglement between $A$ and $B$: Instead, it simply
denotes classical correlation between photon numbers for the two
modes.

Now, however, suppose that some of the remaining laser light is
supplied to Alice (as for instance for the purpose of producing a
local oscillator \cite{akira}).  The overall quantum state between
Alice and Bob will then be of the form
\begin{equation}\label{t2}
\int \frac{{\rm d}\varphi}{2\pi} |T^{\sa{AB}}_\alpha(\varphi)\rangle
\langle T^{\sa{AB}}_\alpha(\varphi)|^{\otimes M/2} \otimes \big( |\alpha_{\sa{A'}}
e^{i\varphi}\rangle\langle \alpha_{\sa{A'}}
e^{i\varphi}|\big)^{\otimes N},
\end{equation}
where $A'$ indicates the further modes in Alice's possession, and where we
assumed all $M$ (assumed an even number) squeezed copies from the state (\ref{sq2}) 
were split on the beamsplitter but the unsqueezed part was not. Far
from being an unentangled state, this state has every bit as much
entanglement as if the laser were actually a pure coherent source.
It is just that the entanglement is in the form of {\it distillable
entanglement\/} \cite{ent}.

To see this, contemplate Alice doing a complete measurement on the
extra laser light in her lab.  With it, she will reduce the quantum
state of modes $A,B$ to a true two-mode squeezed state. Since these
measurements are local (all measurements are performed on Alice's
modes $A'$), it follows there must be distillable entanglement
between Alice's and Bob's modes.  Although the claim in
\cite{rudolph} that the state (\ref{t}) can be produced locally by
Alice and Bob is quite correct, the state (\ref{t2}) is entangled
and cannot be so produced.

For a phase diffusing laser the distillable entanglement in a state
like (\ref{t2}) will be slightly less than the entanglement present
in the corresponding two-mode squeezed state, because a measurement
by Alice on part of her light would reduce the state of modes $A$ and
$B$ to a slightly noisy version of a two-mode squeezed state. 
In the same simple picture as used before, that noise can be understood
as arising from the randomness of the phase $\phi$ of the state
$|T_{\alpha}^{\sa{AB}}(\phi)\rangle$ caused by phase diffusion.

\subsection{Teleportation with Continuous Variables}
Since the state (\ref{t2}) does possess entanglement teleportation
of continuous variables is possible even with lasers in mixed
states. The actual procedure used in \cite{akira} required, as was
noted in \cite{rudolph}, both Alice and Bob to use some of the light
of the same laser that generated the two-mode squeezed state to
perform homodyne detection. The fact that Bob shares laser light
with Alice does not imply however, that they share an active quantum
channel over and above their original entanglement.  One can imagine
that all the light in Alice and Bob's possession (both the shared
two-mode squeezed state and the light for their local oscillators)
was sent to them {\em before\/} any actual teleportation takes place.

This may, if one wishes, be considered an additional shared resource
that had not been made explicit before, but in that regard it is
fairly innocent.  As pointed out in \cite{enk}, such a shared
resource is necessary for any teleportation protocol, irrespective
of its physical implementation. For teleportation with continuous
variables, Alice and Bob need to share a synchronized clock; sharing
some of the laser light is a practical way of implementing this. In fact, 
from a technological point of view laser light gives us
the best possible clock \cite{wiseman}. (Similarly, in order to test Bell 
inequality violations with continuous-variable entangled states the local
oscillator field necessary for the required measurements must be transported
as well, see for example \cite{reid}.) 
Thus, in contrast to \cite{rudolph}, we do not consider the presence
of this resource, which acts as a phase reference, as invalidating
teleportation. An independent party, Victor, who would
like to verify Alice and Bob's teleportation skills, could use his
own laser but has to ``phase-lock'' it (in the sense of Section A)
with Alice's laser. After all, Alice's claim is only that she can
teleport a quantum state of a particular mode: Victor is free to
choose the state to be teleported, but not the Hilbert space.

Finally, a crucial point is that the teleportation procedure as a
whole does not depend on the precise value of the absolute phase
$\varphi$. Therefore, for teleportation to succeed, Alice does not
even have to do an absolute phase measurement to actually distill
the entanglement present in the state (\ref{t2}). Teleportation can
be achieved without knowing the imagined ``unknown'' phase $\varphi$
arising in any PEF\@. Note in particular that Alice and Bob can
teleport a quantum state handed to them by the independent third
party Victor even if he is able to generate a pure coherent state or
a pure entangled state. This is because the phases of both input and
output state are compared to one and the same phase reference.

\section{Conclusions}
In conclusion, viewing the laser beam of a CW laser as a sequence of
$N$ quantum systems led us to the following result: The quantum state
of an ideal laser beam is a mixture of $N$ copies of identical pure
{\it coherent\/} states. By the quantum de Finetti representation
theorem, the coherent states play a unique role in that regard. Such
a state is very different from $N$ copies of identical mixed states
(be they mixtures of number states or of coherent states). One
consequence is that appropriate measurements performed on part of a
laser beam will reduce the quantum state of the rest of the laser
beam to a pure coherent state (or a slightly noisy version thereof
if one considers a realistic laser). Such measurements may in fact
be possible with present-day technology \cite{jones}, and thus an
optical coherent state may in fact be generated.  No sophisticated
measurement on the laser medium \cite{molmer} need be contemplated
to carry this out.

Most importantly, this description allows us to properly assess
quantum communication protocols that rely on lasers. In particular
we found that teleportation with continuous variables is possible
with conventional lasers without actually having to reduce the
quantum state of a laser to a coherent state.

\section*{Acknowledgments}
We thank Terry Rudolph and Barry Sanders for graciously sending us an
early version of their paper, Klaus M\o lmer, Carlton Caves and Ivan
Deutsch for useful comments and Julio Gea-Banacloche for suggesting
the example of Sec.~\ref{atom}.


\begin{thebibliography}{99}

\bibitem{lamb}
M.~Sargent, M.~O. Scully, and W.~E. Lamb, {\em Laser Physics},
Addison-Wesley, Reading (1974).

\bibitem{walls}
D.~F. Walls and G.~J. Milburn, {\em Quantum Optics}, Springer-Verlag,
Berlin 1994.

\bibitem{CavesFuchs96}
C.~M. Caves and C.~A. Fuchs, ``Quantum Information:\ How Much
Information in a State Vector?,'' in {\sl The Dilemma of Einstein,
Podolsky and Rosen --- 60 Years Later}, edited by A.~Mann and
M.~Revzen [Ann.\ Isr.\ Phys.\ Soc.\ {\bf 12}, 226 (1996).

\bibitem{Braunstein99}
S.~L. Braunstein, C.~M. Caves, R.~Jozsa, N.~Linden, S.~Popescu, and
R.~Schack, Phys.\ Rev.\ Lett.\ {\bf 83}, 1054 (1999).

\bibitem{Kok00}
P.~Kok and S.~L. Braunstein, Phys.\ Rev.\ A {\bf 61}, 042304 (2000).

\bibitem{molmer}
K.~M\o lmer, Phys.\ Rev.\ A {\bf 55}, 3195 (1997); J. Mod.\ Optics
{\bf 44}, 1937 (1997).

\bibitem{rudolph}
T.~Rudolph and B.~C. Sanders, Phys. Rev. Lett. {\bf 87}, 077903
(2001).

\bibitem{collett}
M.~J. Collett and C.~W. Gardiner, Phys.\ Rev.\ A {\bf 30}, 1386
(1984).

\bibitem{hudson}
R.~L. Hudson and G.~R. Moody, Z. Wahrs.\ {\bf 33}, 343 (1976).

\bibitem{caves}
C.~M. Caves, C.~A. Fuchs, and R.~Schack, {\tt quant-ph/ 0104088}.

\bibitem{vogel}
W.~Vogel and D.-G. Welsch, {\em Lectures on Quantum Optics},
(Akademie verlag GmbH, Berlin, 1994), Chapter 8.

\bibitem{loudon}
K.~J. Blow, R.~Loudon, S.~J.~D. Phoenix, and T.~J. Shepherd, Phys.\
Rev.\ A {\bf 42}, 4102 (1990).

\bibitem{schack}
R.~Schack, T.~A. Brun, and C.~M. Caves, Phys.\ Rev. A {\bf 64},
6401305 (2001).

\bibitem{drummond}
P.~D. Drummond and M.~D. Reid, Phys. Rev. A {\bf 37}, 1806 (1988).

\bibitem{enkkimble}
S.~J. van Enk and H.~J. Kimble, {\tt quant-ph/0107088}, to be
published in QIC.

\bibitem{noteD}
For this particular problem of two $\pi/2$ pulses
one may, alternatively, define two nonoverlapping functions
$\Phi_1(t)$ and $\Phi_2(t)$ corresponding to the two envelopes of
the pulses that can then be extended to form a complete set of
functions $\{\Phi_i(t)\}$. The two temporal modes thus defined are
then, to a good approximation, the only two nonempty modes. One
finds, as before, a fixed phase relationship between the two
states.  See \cite{enkkimble} for more details.

\bibitem{JulioThing}
It is noteworthy that a very different approach based on so-called
robust unravelings of Master equations describing open quantum
systems also indicates a privileged role for the coherent states,
namely that of the pure states that remain closest to pure during
evolution. See, for example, J.~Gea-Banacloche, Found. Phys. {\bf
28}, 531 (1998); H.~M. Wiseman, Phys. Rev. A {\bf 47}, 5180 (1993).

\bibitem{jones}
D.~J. Jones, S.~A. Diddams, J.~K. Ranka, A.~Stentz, R.~S. Windeler,
J.~L. Hall, and S.~T. Cundiff, Science {\bf 288}, 635 (2000).

\bibitem{bea}
J.~Gea-Banacloche, private communication.

\bibitem{akira}
A.~Furusawa, J.~L. S\o rensen, S.~L. Braunstein, C.~A. Fuchs, H.~J.
Kimble, and E.~S. Polzik, Science {\bf 282}, 706 (1998).

\bibitem{ent}
C.~H. Bennett, D.~P. DiVincenzo, J.~A. Smolin, and W.~K. Wootters,
Phys.\ Rev.\ A {\bf 54}, 3824 (1996).

\bibitem{enk}
S.~J. van Enk, J.\ Mod.\ Optics {\bf 48}, 2049 (2001).

\bibitem{wiseman}H.~M. Wiseman, {\tt quant-ph/0104004}.

\bibitem{reid}M.~D. Reid, Phys. Rev. Lett. {\bf 84}, 2765 (2000) 
and references therein.

\end{thebibliography}
\end{document}